\documentstyle[11pt,paspconf,epsf]{article}

\begin{document}

\title{The Gaseous ISM: Observations with the Wisconsin H$\alpha$ Mapper
(WHAM)}
\author{R. J. Reynolds, L. M. Haffner, and S. L. Tufte}
\affil{Astronomy Department, University of Wisconsin, Madison, WI 53706}

\begin{abstract}

The Wisconsin H$\alpha$ Mapper (WHAM) is a new facility dedicated to the
study of faint optical emission lines from diffuse interstellar gas.  
During its first 18 months of operation, WHAM carried out a survey of the
interstellar H$\alpha$ emission associated with the warm, ionized
component of the interstellar medium.  The observations consisted of
37,000 spectra obtained with a one degree diameter beam on a 0${\rm^o}$.98
$\times$ 0${\rm^o}$.85 grid ($l \times b$), covering the sky above
declination --30${\rm^o}$. This survey provides for the first time a
detailed picture of the distribution and kinematics of the diffuse ionized
hydrogen through the H$\alpha$ line comparable to surveys of the neutral
hydrogen obtained through the 21 cm line.  Preliminary reduction of the
data from selected portions of the sky reveal that the interstellar H~II
has a complex distribution, with long filaments and loop-like structures
extending to high Galactic latitudes and superposed on a more diffuse
background.  Apart from the H$\alpha$ sky survey, WHAM also has detected
for the first time faint diagnostic emission lines in selected directions,
[O~I] $\lambda$6300, [O~III] $\lambda$5007, and He~I $\lambda$5876, which
provide information about the physical state of the gas and clues
about the source of the ionization.  Maps of [S~II] $\lambda$6716 and
[N~II] $\lambda$6584 over limited regions of the sky are providing
information about variations in the temperature and ionization conditions
within the Galactic disk, and the detection of faint optical emission
lines from high velocity clouds is probing conditions in the halo.  
Finally, WHAM has the capability to explore the smaller scale structure of
the medium through very narrow band (12 km s$^{-1}$), 1$^{\prime}$ angular
resolution images within selected 1${\rm^o}$ fields.  This facility is
currently located at Kitt Peak National Observatory near Tucson, Arizona
and operated remotely from Madison, Wisconsin.
 
\end{abstract}


\section{Properties of the Warm Ionized Medium}

     Diffuse ionized gas is a major, yet little understood component of
the interstellar medium, which consists of regions of warm (10$^{4}$ K),  
low-density (10$^{-1}$ cm$^{-3}$), nearly fully ionized hydrogen that
occupy approximately 20\% of the volume within a 2 kpc thick layer about
the Galactic midplane (e.g., Reynolds 1991, 1993a).  Hoyle \&
Ellis (1963) were the first to suggest the existence of an extensive layer
of warm H II surrounding the Galactic disk, based on the spatial and
spectral characteristics of the Galactic synchrotron background at very   
low radio frequencies (2-10 MHz).  However, it was not until the discovery
of pulsars a few years later that the existence of widespread H II in the
diffuse interstellar medium became generally accepted (see review by   
Gu\'elin 1974).  This was soon followed by the detection of faint,
diffuse, optical line emission from the Galaxy (Reynolds, Roesler, \&
Scherb 1974), which provided information about the temperature, ionization
state, kinematics, spatial distribution, and power requirements of the gas
within 2--3 kpc of the sun.  More recent narrow band filter imaging has
revealed that widespread ionized hydrogen is also
present in other galaxies (e.g., Rand, Kulkarni, and Hester 1990; Dettmar
1992; Hunter and Gallagher 1990; Walterbos and Braun 1994; Ferguson et al
1996; and Rossa and Dettmar at this meeting).

At the Galactic midplane, the space averaged density of H~II is less than
5\% that of the H I.  However, because of its greater scale height, the
total column density of interstellar H II along high Galactic latitude
sight lines is relatively large, 1/4 to 1/2 that of the H I, and one
kiloparsec above the midplane, warm H~II may be the dominant state of the
interstellar medium (Reynolds 1991).  The existence of this
ionized medium can have a significant effect upon the interstellar
pressure near the Galactic midplane (Cox 1989) and upon the dynamics of
hot (10$^5$ -- 10$^6$ K), ``coronal'' gas far above the midplane (e.g.,
Heiles 1990).

     Neither the source of the ionization nor the relationship of this gas
to the other components of the medium, such as the H I, is understood.   
The intensity of the H$\alpha$ background at high Galactic latitude
implies an average of $5 \times 10^6$ H-ionizations s$^{-1}$ per cm$^2$
column perpendicular to the Galactic disk (Reynolds 1992), which  
corresponds to a power input of at least $1 \times 10^{-4}$ ergs s$^{-1}$
cm$^{-2}$ (at 13.6 eV per ionization).  Of the known sources of ionization
within the Galaxy, only O stars, which produce $3 \times 10^7$ ionizing   
photons s$^{-1}$ cm$^{-2}$ (e.g., Abbott 1982; Vacca et al 1996)
comfortably exceed this requirement.  Supernovae, which inject
approximately $1 \times 10^{-4}$ ergs s$^{-1}$ cm$^{-2}$ of kinetic energy
into the interstellar medium, cannot be the primary source unless the
ionization process is nearly 100\% efficient, or unless supernovae in the
Galaxy are significantly more numerous and energetic than the generally   
assumed Galactic values of $5 \times 10^{50}$ ergs per 40 years.  Models 
by Miller \& Cox (1993) and Dove \& Shull (1993) have shown that for
certain distributions of the H I clouds and intercloud medium, Lyman   
continuum photons from O stars and OB associations could travel far from  
the midplane, ionizing large volumes of interstellar space.  Therefore, it
is tempting to conclude that O stars must be the source, and that the
regions of diffuse H II are very extended ``Str\"omgren Spheres'' in a
diffuse intercloud medium as first suggested by Davidson \& Terzian
(1969).

     However, if O stars are to account for the ionization, then their
photons must travel not only large distances above the midplane, but also
within the H I cloud layer, which would require significantly fewer H I
clouds per kpc than the ``standard model'', at least along the one
sightline that was examined (Miller \& Cox 1993; also Reynolds 1993b).  
Furthermore, the ionization conditions within the diffuse ionized gas
appear to differ significantly from conditions within O star H II
regions.  The anomalously strong [S~II] $\lambda$6716/ H$\alpha$ and weak
[O~III] $\lambda$5007/ H$\alpha$ emission line ratios (compared to O and B
star H II regions) indicate a low state of excitation with few ions
present that require ionization energies greater than 23 eV (Reynolds
1985b, 1988).  This is supported by the low ionization fraction of helium
in the diffuse H II (Tufte 1997; Reynolds \& Tufte 1995; Heiles et al
1996), which implies that the spectrum of the diffuse interstellar
radiation field that ionizes the hydrogen is significantly softer than
that from the Galactic O star population.  The combination of high [S II]
intensities, low [O I] $\lambda$6300 intensites (Reynolds et al 1998b;
Dettmar \& Schulz 1992), and low He I intensities relative to H$\alpha$
are not accounted for simultaneously by current O star photoionization
models (e.g., Domg\"{o}rgen \& Mathis 1994; Sokolowski 1994).  The data
seem to require either a softening of the radiation after it leaves the
vicinity of the O star, with an additional source of heating within the
diffuse ionized regions, or an ionizing source other than O stars
(see Rand 1997, 1998).  Suggested sources other than O stars have
included, for example, galactic flares (Raymond 1992), turbulent mixing
between the hot and warm phases (e.g., Slavin, Shull, \& Begelman 1993),
cosmic ray electrons (Ramaty \& Skibo 1993), and the decay of dark matter
(Sciama 1993).

\section{The Wisconsin H$\alpha$ Mapper Facility}

     The Wisconsin H$\alpha$ Mapper (WHAM) is a recently completed
facility (funded by the National Science Foundation) for the detection and
study of faint emission lines from diffuse ionized gas in the disk and
halo of the Galaxy. WHAM consists of a 15 cm aperture, dual etalon
Fabry-Perot spectrometer coupled to a dedicated 0.6 m light bucket
telescope, which provide a one degree diameter beam on the sky and produce
a 12 km s$^{-1}$ resolution spectrum across a 200 km s$^{-1}$ spectral
window (Reynolds et al 1990; Tufte 1997).  The spectral window can be set
to any wavelength between 4800 \AA\ and 7200 \AA\ using a gas 
pressure control system to tune the Fabry-Perot etalons and a filter wheel
to provide the correct isolating interference filter.  The tandem etalons
greatly extend the effective ``free spectral range'' of the spectrometer
and suppress the multi-order Fabry-Perot ghosts, especially those arising
from the relatively bright atmospheric OH emission lines within the pass
band of the interference filter.  A high quantum efficiency (78\% at
H$\alpha$), low noise ( 3 e$^{-}$ rms) CCD camera serves as a multichannel
detector, recording the spectrum as a Fabry-Perot ``ring image'' (Reynolds
et al 1990).  The combination of high spectral resolution, the largest
available Fabry-Perot etalons, and a high efficiency CCD on a dedicated
telescope make WHAM a powerful instrument for exploring many aspects of
the the warm ionized medium.

     The WHAM facility is located on Kitt Peak near Tucson Arizona.  The
siderostat, the CCD camera and LN$_2$ camera filling system,
the etalon pressure system, the interference filter wheel, the calibration
light sources and their flip mirror, plus a number of environmental
sensors can provide information to and can be controlled by a single
workstation.  By incorporating the messaging system developed for the WIYN
telescope (Percival 1994), the entire WHAM facility, including opening and
closing the telescope, can be operated from a remote location.  This
capability allows WHAM to have the benefits of a clear air site, operating
about 70 clear, dark of the moon nights per year with relatively little
travel for the observers.  Normal operation requires a part-time, on-site
technician/observer for minor maintenance, trouble shooting, and
monitoring sky conditions, plus an occasional trip from Wisconsin for
special calibrations, and changes to special filters not on
the wheel.

\section{Observations}

     A number of new investigations have been started with this facility,
including the H$\alpha$ sky survey, mapping portions of the sky in
forbidden lines, studies of optical emission lines from high velocity
clouds and cloud complexes, the detection of extremely faint emission
lines that probe physical conditions within the gas, and high angular
resolution (1$^{\prime}$), high velocity resolution (12 km s$^{-1}$)
imaging.

\subsection{The H$\alpha$ Sky Survey}

     Between January 1997 and September 1998, WHAM carried out one of its
primary missions, a northern sky H$\alpha$ survey of the warm ionized
medium.  This survey consists of approximately 37,000 spectra above
declination $-30^{\rm o}$, sampling the sky on a 0$^{\rm o}$.85 $\times$
0$^{\rm o}$.98 grid with a 1$^{\rm o}$ beam.  Each spectrum has a 30 s
integration time and covers a 4.4 \AA\ (200 km s$^{-1}$) spectral interval
centered near the LSR at a resolution of 0.26 \AA\ (12 km s$^{-1}$).  All
observations were carried out during dark of the moon to avoid
contamination by features in the solar spectrum.  This survey has provided
the first detailed view of the distribution and kinematics of the diffuse
$ionized$ hydrogen through the optical H$\alpha$ line comparable to the
large scale survey maps of the $neutral$ hydrogen obtained through the
radio 21 cm line.

     Figure 1 illustrates a sample spectrum from the survey at
\begin{figure}
  \vspace{0.125in}
  \centerline{\hbox{
      \epsfysize=1.65in \epsfbox{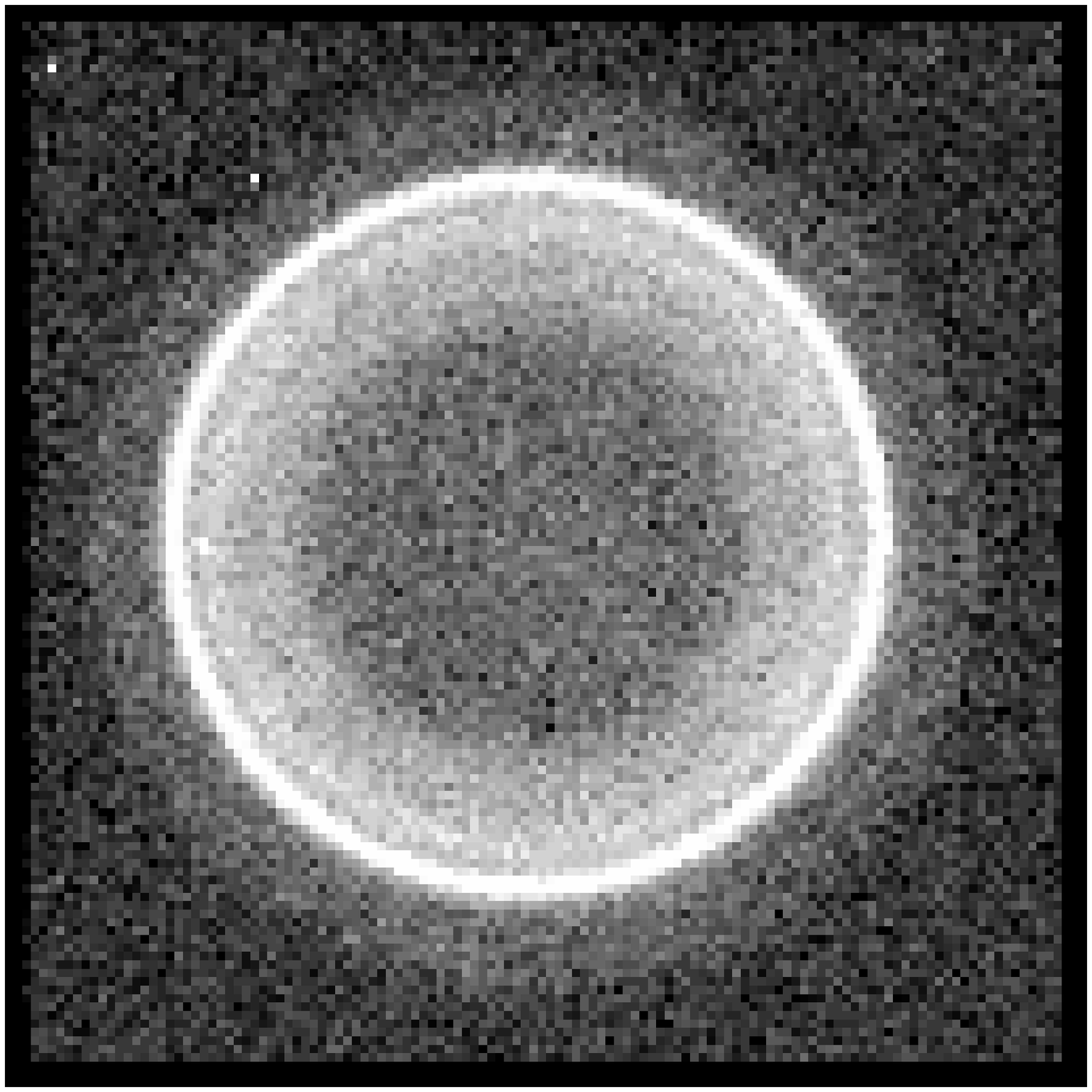}
      \epsfysize=1.77in \epsfbox{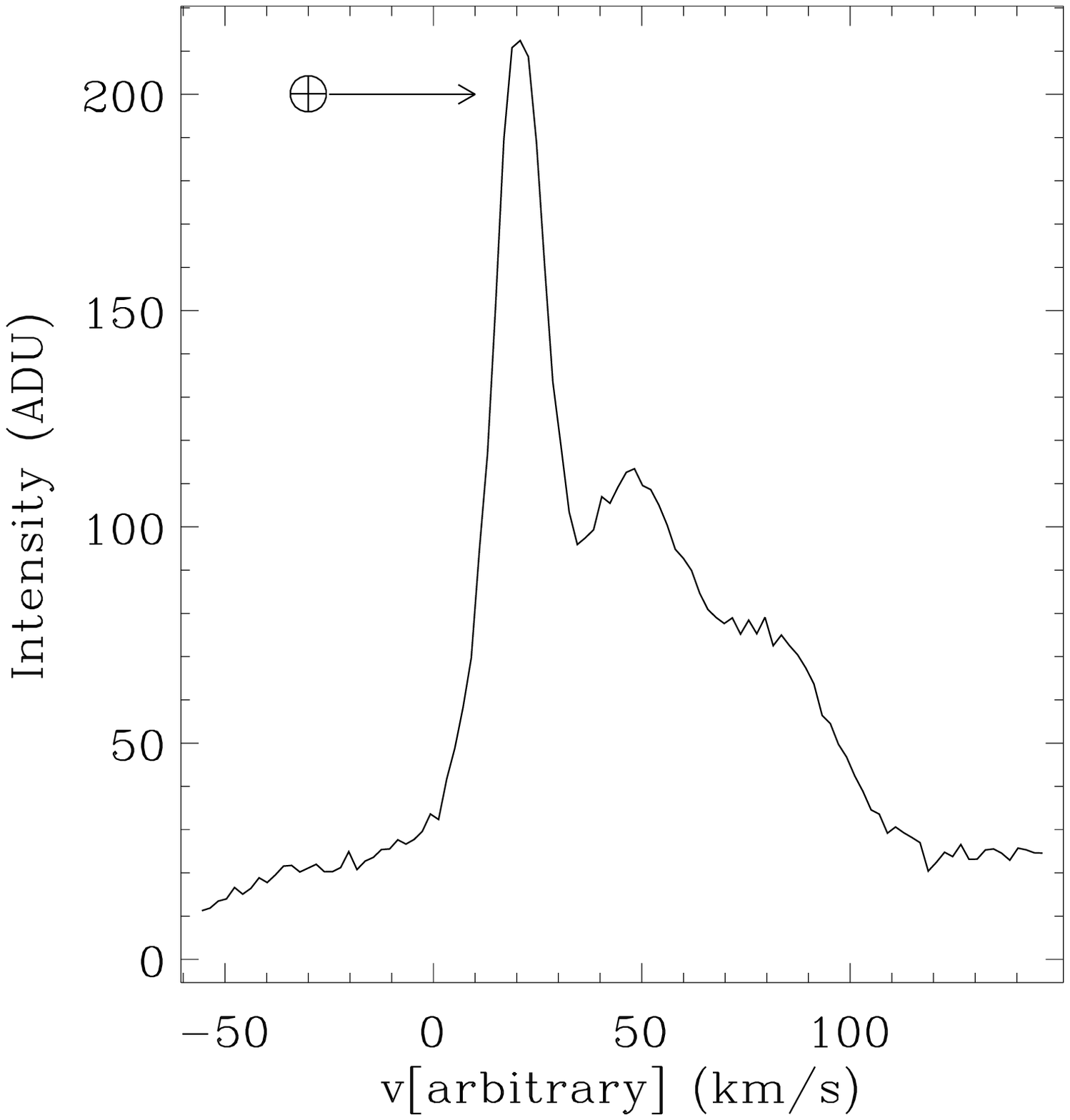}
      \epsfysize=1.77in \epsfbox{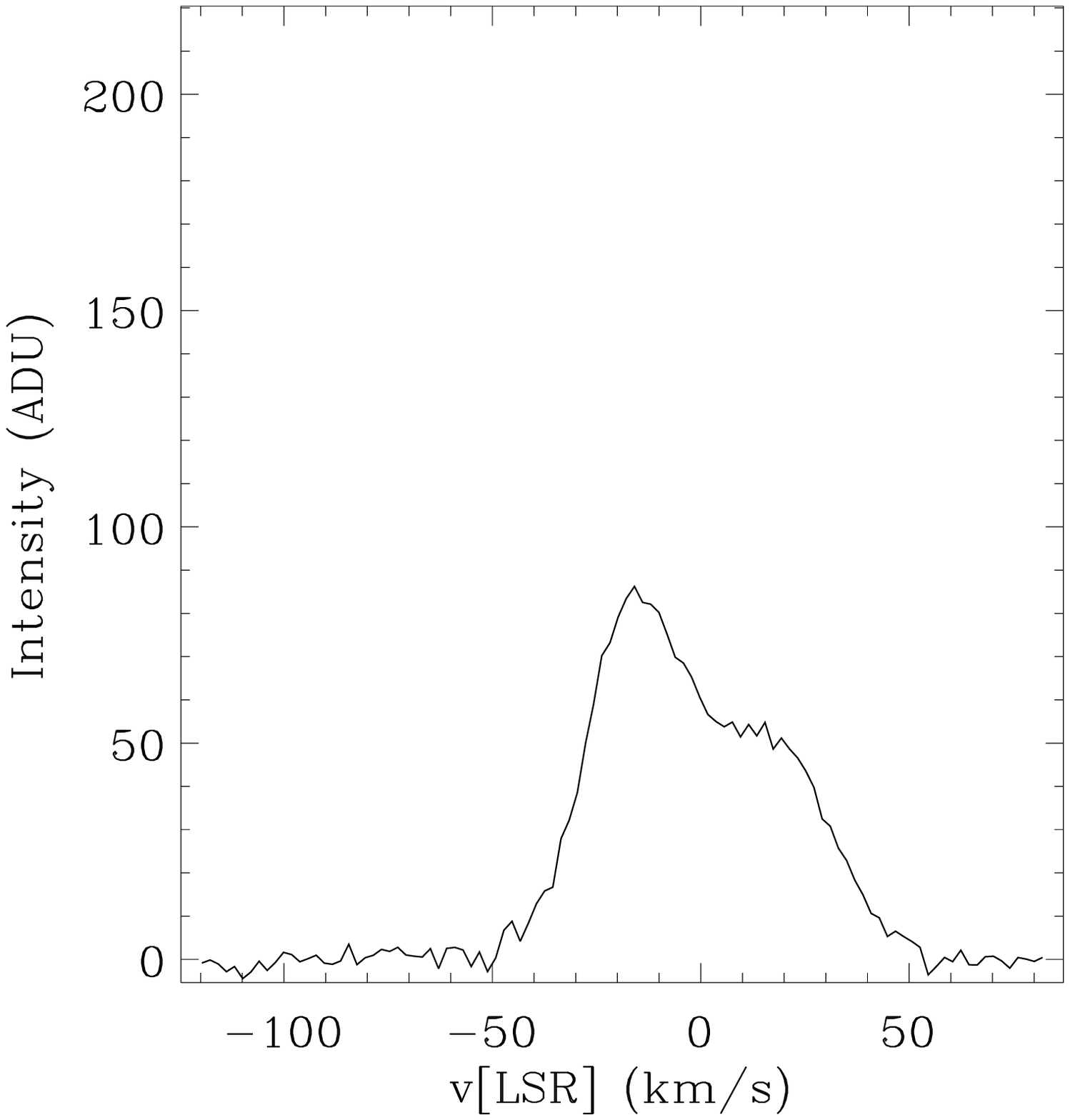}}
    }
\caption{WHAM H$\alpha$ data for $l = 203.^{\rm o}8, b = -43.^{\rm   
o}3$. a) the raw CCD image (30 s exposure); b) the resulting spectrum
($\oplus$ denotes the geocoronal line); c) the pure interstellar spectrum
after flat fielding, removal of the geocoronal line, subtraction of       
the sky continuum, and registration of the velocity scale to the LSR.}
\end{figure}
$l = 203.^{\rm o}8$, $b = -43.^{\rm o}3$, showing both the raw CCD image
and resulting H$\alpha$ spectra.  The geocoronal H$\alpha$ line, produced
by solar excitation of atomic hydrogen in the earth's upper atmosphere, is
the thin, bright annulus in the CCD ``ring spectrum''.  This emission
appears as a prominent, relatively narrow line in the center frame of
Figure 1.  The interstellar emission is the broader feature inside the
geocoronal ring, appearing in this case to consist of two blended velocity
components at +30 km s$^{-1}$ and +60 km s$^{-1}$ with respect to the
geocoronal line. In general, the separation between the interstellar
emission and the geocoronal line is due to a combination of the earth's
orbital velocity, the sun's peculiar velocity, and intrinsic motions of
the interstellar gas, including Galactic differential rotation.  The two
interstellar components have intensities of about 2 R and 4 R, where a
Rayleigh, 1 R = 2.41 $\times 10^{-7}$ erg cm$^{-2}$ s$^{-1}$ sr$^{-1}$ at
H$\alpha$ and corresponds to an emission measure of 2.3 cm$^{-6}$ pc for a
temperature of 8000 K.  The geocoronal line is removed from the data by
fitting each spectrum with gaussian components and then subtracting from
the spectrum the fitted gaussian associated with the geocorona.  The
resulting interstellar spectrum is shown in the third frame of
Figure 1.  The absolute intensity calibration is obtained by
comparison with standard astronomical sources (e.g., Scherb 1981).

     A portion of the survey data is presented in Figure 2 as a gray scale
\begin{figure}
  \centerline{\epsfysize=3.5in \epsfbox{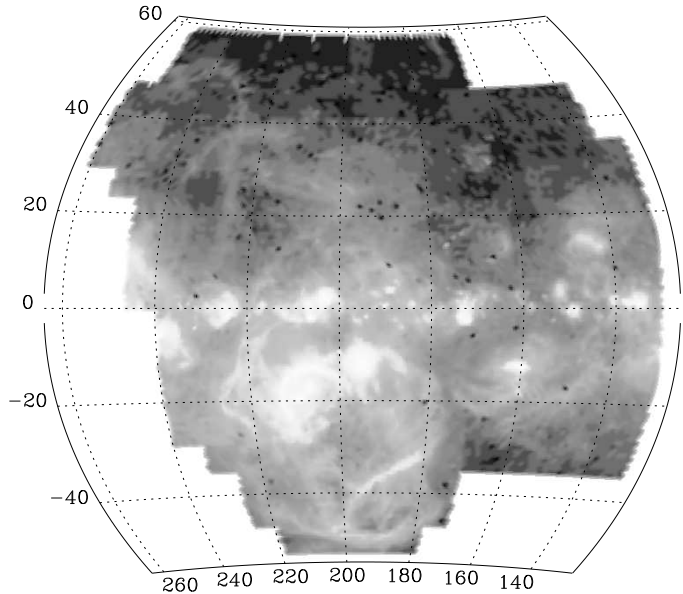}}
  \caption{Total intensity map of the diffuse interstellar H$\alpha$
background extracted from a portion of the WHAM sky survey.  The map is in
Galactic coordinates, with the Galactic equator running horizontally
through the center.  The map is bounded on the left by the $-30^{\rm o}$
declination limit of the survey. The small black dots are pixels 
contaminated by a bright star within the beam.}
\end{figure}
map of the total intensity of the interstellar H$\alpha$ emission. This
map, synthesized from approximately 13,000 H$\alpha$ spectra, covers the
region of the sky between about 130$^{\rm o}$ to 240$^{\rm o}$ Galactic
longitude and $\pm$50$^{\rm o}$ Galactic latitude.  The gray scaling and
stretch have been adjusted to reveal the fainter high latitude emission
along with the brighter regions near the plane.  Interstellar H$\alpha$
emission is detected in every direction, with intensities that range from
thousands of Rayleighs near the Orion nebula ($\sim$ 209$^{\rm
o}$,$-20^{\rm o}$) and $\sim$ 100--200 R in Barnard's loop and the large
$\lambda$ Ori H II region (195$^{\rm o}$, $-13^{\rm o}$) to $\sim$ 0.5 R
in some of the fainter high latitude regions (e.g., 220$^{\rm o}$,
+45$^{\rm o}$).  The map reveals numerous large scale filaments superposed
on a fainter H$\alpha$ background.  A number of ``classical'' H II regions
also dot the map near the Galactic equator. Some of the filamentary
features are associated with the Orion-Eridanus bubble, which fills the
sky from $l$ = 180$^{\rm o}$ to 210$^{\rm o}$ and $b = -10^{\rm o}$ to
$-50^{\rm o}$ and includes Barnard's Loop near its northern boundary.  
Many of the Orion-Eridanus filaments are correlated with emission features
at 21 cm and x-ray wavelengths (see, for example, Reynolds \& Ogden 1979;
Brown, Hartmann, \& Burton 1995; Burrows et al 1993; Heiles, this
meeting).  However, other filaments on the map have no obvious
correspondence to any previously known structures, for example, the 1 R
feature rising vertically from 226$^{\rm o}$, +10$^{\rm o}$.

     This latter filament is approximately 2$^{\rm o}$ wide and at least
80$^{\rm o}$ long, terminating at the southern boundary of the survey, $l$
= 270$^{\rm o}$, $b$ = +42$^{\rm o}$ (see Haffner, Reynolds, and Tufte
1998). The vertical portion of this feature between $b = +10^{\rm o}$ and
+25$^{\rm o}$ is associated with a single radial velocity component
centered at +16 km s$^{-1}$ (LSR).  As the feature arches toward higher
longitudes the velocity decreases, reaching about --20 km s$^{-1}$ at the
southern limit of the survey.  Where the filament appears to meet the
Galactic plane near $l$ = 225$^{\rm o}$, it is directly above the H II
region surrounding the CMa OB1 association, which has a radial velocity
similar to that of the filament just above it.  The coincidence in
location and radial velocity of one end of the filament with this
energetic source in the Galactic midplane suggests a possible connection
between them.  However, the nature of the connection is not clear, because
the photoionizing flux from the O stars in the association and the
recombination and kinematic time scales for the gas in the filament appear
to be inconsistent with ejection or ionization by the association (see
discussion in Haffner et al 1998).  Although the origin of this and other
filaments revealed by the survey has not yet been identified, their
existence provides important new clues about the nature of the ionized
medium at high latitudes.

     Once the reduction of the survey data is completed (mid 1999), it
will be possible to carry out for the first time a detailed comparison of
the distribution and kinematics of the H II with that of the H I.  One
question such a comparison may help to address is the basic structure of
the interstellar medium: the distribution and topology of clouds, the warm
``intercloud'' medium, and the hot coronal gas.  For example, it is not
yet know whether the warm component (the warm neutral medium and the warm
ionized medium) is confined to isolated clouds embedded within a pervasive
hot medium (McKee \& Ostriker 1977) or whether the warm component is in
fact the pervasive medium with the hot gas confined to relatively small,
isolated bubbles within it (Heiles 1990; Slavin \& Cox 1993).  The
distribution of the warm diffuse H II and its spatial and kinematic
relationship to the H I has a direct bearing on this question. In the
McKee \& Ostriker model, the warm H II is located in the envelopes of H~I
clouds and therefore is a tracer of the $boundaries$ between the warm and
the hot, ``coronal'' phases.  On the other hand, if warm intercloud gas is
the pervasive medium, then the H~II will have a fundamentally different
distribution, occupying extensive regions $between$ the H I clouds (e.g.,
Miller \& Cox 1993).

     In addition to the 21 cm studies, this H$\alpha$ survey also may
impact the interpretation of other observations, including:

\begin{itemize}

\item the cosmic microwave background, which is contaminated by the
free-free continuum emission associated with the diffuse H II (e.g.,
Gundersen et al 1996; Bennett et al 1992);

\item the far-UV background, which has a significant ($\sim$ 10\% -- 20\%)
contribution from the hydrogen two-photon continuum emitted by the diffuse
H II (Reynolds 1992);

\item the 100 MeV diffuse $\gamma$ -ray background, produced by the
interaction of cosmic rays with interstellar matter, including the H
II (Bloemen 1989), which is 0.26 -- 0.63 of the H I column density at high
latitudes;

\item the far-IR emission line background such as [N II] $\lambda$205
$\mu$m and [C II] $\lambda$158 $\mu$m (Petuchowski \& Bennett 1993; Heiles
1994);

\item UV absorption line studies, including the origin and location of   
moderate to low ions such as S$^+$, C$^+$, Al$^{++}$, etc. (e.g., Spitzer  
\& Fitzpatrick 1993; Savage \& Howk 1998);

\item rotation measures, radio scintillation, and angular broadening
observations (e.g., Minter \& Spangler 1996);

\item pulsar dispersion measures and distances (Taylor \& Cordes 1993);
and

\item other, complementary H$\alpha$ surveys that map the structure of
the ionized gas at higher angular resolution, but lack velocity resolution
and an absolute intensity scale (e.g., Gaustad et al 1996; Simonetti et al
1996; Parker \& Phillipps 1998).

\end{itemize}

\subsection{Maps of [S II] $\lambda$6716 and [N II] $\lambda$6584}

     Another program involves mapping limited regions of the sky in
emission lines other than H$\alpha$.  The forbidden lines, [S II]
$\lambda$6716 and [N II] $\lambda$6584 are the two brightest optical lines
in the interstellar background after H$\alpha$, with intensities typically
about 40\% that of H$\alpha$, and they are a probe of physical
conditions within the ionized medium.  The [S II] line, for example, is
about five times brighter relative to H$\alpha$ in the diffuse
interstellar medium than in traditional H II regions surrounding O and B
stars (e.g., Haffner 1998; Reynolds 1985b, 1988).  Elevated [S
II]/H$\alpha$ (and [N II]/H$\alpha$) intensity ratios have also been found
to be a characteristic of the warm ionized medium in other galaxies (e.g.,
Walterbos \& Braun 1992), with ratios increasing as a function of distance
from the midplane (Dettmar \& Schulz 1992; Golla, Dettmar, \& Domg\"orgen
1996; Rand 1997).  In the Milky Way there also are real variations in the
[S II]/ H$\alpha$ ratio, ranging from about 0.2 to 1.0 from sightline to
sightline (Haffner 1998).  Even along a single sightline the [S
II]/H$\alpha$ ratio can vary greatly from one radial velocity component to
the next.  A preliminary map of the [S II]/H$\alpha$ ratio in a 1000
square degree region bounded by $124^{\rm o} < l < 163^{\rm o}, -34^{\rm
o} < b < -6^{\rm o}$ reveals large scale variations in the ratio that are
anticorrelated with the H$\alpha$ surface brightness (Haffner 1998).

     The reason for the anomalously high line ratios and their variations
is not well understood, but they are potentially important clues to the
excitation and ionization conditions within the gas. Photoionization
models (Domg\"{o}rgen \& Mathis 1994; and Sokolowski 1994) indicate that a
low ionization parameter, the ratio of photon density to gas density, is
probably one of the principal factors producing elevated [S II]/ H$\alpha$
and [N II] /H$\alpha$ ratios.  However, the models are not successful at
explaining the low [O I] and He I intensities (see below), and, therefore,
other factors such as electron temperature and the spectrum of the
radiation field may also be important (Rand 1997, 1998).  Because nitrogen
and sulfur have different ionization potentials, they respond differently
to changes in ionization conditions.  On the other hand, because the two
emission lines, [N II] $\lambda$6584 and [S II] $\lambda$6716, are
associated with metastable states that have nearly the same energy above
the ground state, their intensities should respond nearly identically to
variations in electron temperature.  Therefore, maps of [S II] and [N II]
in addition to the H$\alpha$, in combination with photoionization models,
can probe ionization and excitation conditions and their variations
throughout the disk and halo.

\subsection{Optical Emission Line Studies of High Velocity Clouds}

     Although their existence has been known for many years, the origin of
High Velocity Clouds (HVCs) is still a mystery (e.g., Wakker et al. 1996).
This is due at least in part to the fact that until relatively recently,
HVCs could only be studied via the 21 cm emission line.  Absorption line
observations have now begun to place significant constraints on the
distance and metallicity of some of the clouds (e.g., Danly, Albert, \&   
Kuntz 1993; Wakker et al 1996, 1998), and there has even been evidence
recently presented for associated soft x-ray emission (Herbstmeier et al.
1995).

     The detection of HVCs in optical emission at H$\alpha$ (Kutyrev \&
Reynolds 1989) and, with WHAM, in [S II] $\lambda$6716 in addition to
H$\alpha$ (Tufte et al 1998) have opened a new window through which to
explore the nature of these objects.  Initial studies of the M I cloud
with WHAM have shown, for example, that the H$\alpha$ emission, while
clearly associated spatially and kinematically with the 21 cm emission,
does not correlate closely with the H~I column density.  In the M I cloud
there is some indication that the ionized gas producing the H$\alpha$
emission envelops the neutral gas; however, the ``H II halo'', if present,
is not large.  Furthermore, the [S II]/H$\alpha$ ratio varies
significantly, indicating large variations in the ionization/excitation
conditions within the cloud.  The line widths imply a temperature near
10$^4$ K and low ($\leq 20$ km s$^{-1}$) nonthermal velocities, suggesting
photoionization rather than shock excitation.  Finally, because at least
some HVCs are located far from the Galactic midplane (Wakker et al 1996),
the optical line observations provide the opportunity to probe the
environment far outside the Galactic disk (see, for example, discussions
by Weiner \& Williams 1996 and Bland-Hawthorn \& Maloney 1998).

\subsection{Extremely Faint Diagnostic Lines}
     
     WHAM provides the opportunity to study emission lines
that are too faint to have been detected previously.  The intensities,
line widths, and radial velocities of these lines give information on the
physical conditions within the gas and place strong constraints on
theoretical models.  Examples of such diagnostic lines include:

\begin{itemize}

\item [He I] $\lambda$5876 -- This line is interesting because
recombination lines of He I in the diffuse ionized gas of the Galaxy
appear to be significantly fainter relative to those of hydrogen than O
star ionization models predict (Tufte 1997; Reynolds \& Tufte 1995;  
Heiles et al 1996; Domg\"{o}rgen \& Mathis 1994).  He~I in the edge-on
galaxy NGC 891 is stronger than in the Milky Way (Rand 1997, 1998), but
the basic conclusion is the same as for the Milky Way, namely, the
ionizing spectrum appears to be significantly softer than that of the
assumed source spectrum (O stars).  This presents a challenge to
the O star models for the warm ionized medium, especially since the models
seem to require a $harder$ spectrum than O stars to explain the relatively
high forbidden line intensities (Sokolowski 1994; Rand 1998).

\item [[ O I]] $\lambda$6300 -- The ionization fraction of oxygen is tied
closely to that of hydrogen through the large H$^+$ + O$^{\rm o}
\Longleftrightarrow$ H$^{\rm o}$ + O$^+$ charge exchange cross section,
such that the [O I] $\lambda$6300/ H$\alpha$ intensity ratio is an
accurate measure of the hydrogen ionization fraction H$^+$/H within the
warm ionized medium.  Observations of this line, whose intensity relative
to H$\alpha$ ranges from less than 0.01 to about 0.04 (Reynolds et al
1998b; Dettmer \& Schulz 1992), imply that the hydrogen within the
H$\alpha$ emitting regions is nearly fully ionized.  These observations
also provide strong constraints on the photoionization models, forcing the
incorporation of optically thin (to Lyman continuum radiation) clouds and
the escape of a significant fraction of the ionizing radiation from the
galaxy (Domg\"{o}rgen \& Mathis 1994; Sokolowski 1994).

\item [[ O III]] $\lambda$5007 -- At low Galactic latitudes the optical [O
III]/ H$\alpha$ intensity ratio is significantly fainter in the Galactic
background than in traditional H II regions (Reynolds 1985a), consistent
with the apparently low ionization state of the warm ionized medium and
the 35 eV ionization potential of O$^{+}$.  Preliminary observations with
WHAM have detected for the first time the optical [O III] line at high
Galactic latitudes (Haffner et al 1996).  Now that it has been detected,
an important question is whether the emission is associated with the warm
component or with the hotter (10$^5$ K) ``coronal'' gas component of the
interstellar medium.  The [O III] $\lambda$5007 intensities at high
latitude are, in fact, comparable to the $\lambda$5007 intensities that
are predicted from the hotter medium (Haffner et al 1996) based on the
intensities of the O III] $\lambda$1663 emission measured toward six high
latitude sightlines by Martin \& Bowyer (1990) during the Berkeley UVX
Shuttle mission.  The high latitude O III] $\lambda$1663 emission (along
with C ~IV $\lambda$1550) is associated with the 10$^5$ K phase of the
medium (Martin \& Bowyer 1990).  If the optical (5007) line is also
associated with the hot gas, it would provide a unique opportunity to map
the distribution and kinematics of this gas, which is believed to occupy a
transition region between the 10$^4$ K and 10$^6$ K phases.

\end{itemize}

\subsection{Deep, High Angular Resolution Imaging}

In addition to spectra, the WHAM spectrometer can obtain very narrow-band,
deep images at 1$^{\prime}$ -- 2$^{\prime}$ angular resolution within a
1$^{\rm o}$ circular field.  In the normal ``spectrum mode'' only the
spectrum of the source, averaged within the beam, is imaged on the
detector.  This eliminates confusion between spectral features and spatial
features (including stars) within the 1$^{\rm o}$ beam.  However, with the
insertion of additional lenses into the optical path, a monochromatic
image of the sky within the 1$^{\rm o}$ diameter beam is recorded on the
detector.  The spectral width of the image can be adjusted by an iris
diaphram to any value between 10 km s$^{-1}$ (0.2 \AA\ at H$\alpha$) and
200 km s$^{-1}$ (4.4 \AA).  This allows imaging in selected radial
velocity components and even portions of a line profile (e.g., red side vs
blue side), if desired. This imaging mode has been used to study the
distribution and kinematics of ions in comets (e.g., Scherb et al 1996),
but has not yet been applied to observations of the interstellar medium.  
Narrow band imaging of the warm ionized medium, when combined with DRAO 21
cm survey observations, for example, could probe the relationship between
the H II and the H I phases at sub parsec scales.  An earlier (pre WHAM)
comparison of 3$^{\prime}$ resolution H$\alpha$ and 21 cm images in one
field ($l = 140^{\rm o}, b = -22^{\rm o}$) containing a moderate velocity
($-65$ km s$^{-1}$) cloud, revealed a close relationship, both spatially
and kinematically, between the H II and H I in the cloud, but also
indicated that the regions of ionized and neutral gas are spatially
separated (Reynolds et al 1995).

\section{Conclusions}

     High-throughput, high spectral resolution Fabry-Perot spectroscopy of
faint interstellar emission lines at optical wavelengths is providing new
information about the distribution, kinematics, and physical conditions of
the diffuse ionized component of the interstellar medium.  The comparison
of these optical emission line data with observations at other
wavelengths, particularly observations that probe other phases of the
medium, should lead to an improved understanding of the composition and
structure of the interstellar medium and the principal sources of
ionization and heating within the Galactic disk and halo.

\acknowledgments

We thank N. R. Hausen, M. Quigley, K. Jaehnig, J. Percival, and T.
Tilleman for their very important contributions to the WHAM program.
WHAM is funded by the National Science Foundation through grant AST
96-19424.

\end{document}